\documentclass[sigconf,10pt,screen]{acmart}
\AtBeginDocument{%
  }

\usepackage{graphicx}
\usepackage{textcomp}
\usepackage{xcolor,colortbl}
\usepackage{setspace}
\usepackage{textcomp}
\usepackage{multirow}
\usepackage{todonotes} 
\usepackage{booktabs}
\usepackage{subcaption}

\usepackage{xspace}
\newcommand{\kernel}{Fused3S\xspace}

\newcommand{\tcBlock}{TCB\xspace}
\newcommand{\tcBlocks}{TCBs\xspace}
\newcommand{\rowWindow}{RW\xspace}
\newcommand{\rowWindows}{RWs\xspace}

\usepackage{algorithm}
\usepackage{algpseudocodex}
\algrenewcommand\algorithmicindent{1em}

\usepackage[mathscr]{eucal}

\usepackage{stmaryrd}

\usepackage{amsbsy}

\usepackage{bm}

\usepackage{braket}

\usepackage{mathtools}

\usepackage{array}

\usepackage{cleveref}

\usepackage{siunitx}
\newcommand{\ie}{i.e., }
\newcommand{\eg}{e.g., }

\newcommand{\Tra}{{\sf T}}

\newcommand{\V}[2][]{{\bm{#1\mathbf{\MakeLowercase{#2}}}}} 
 
\newcommand{\M}[2][]{{\bm{#1\mathbf{\MakeUppercase{#2}}}}} 
 
\newcommand{\rdot}{\textcolor{red}{$\bullet$}}
\newcommand{\gdot}{\textcolor{green}{$\bullet$}}

\definecolor{darkgreen}{RGB}{0,176,6}
\definecolor{dandelion}{RGB}{200,170,70}

\copyrightyear{2025}
\acmYear{2025}
\setcopyright{cc}
\setcctype{by}
\acmConference[ICS '25]{2025 International Conference on Supercomputing}{June 8--11, 2025}{Salt Lake City, UT, USA}
\acmBooktitle{2025 International Conference on Supercomputing (ICS '25), June 8--11, 2025, Salt Lake City, UT, USA}
\acmDOI{10.1145/3721145.3730430}
\acmISBN{979-8-4007-1537-2/2025/06}

\begin{document}

\title{\kernel: Fast Sparse Attention on Tensor Cores}
\author{Zitong Li}
\email{zitongl5@uci.edu}
\affiliation{%
  \institution{University of California, Irvine}
  \country{}
}
\author{Aparna Chandramowlishwaran}
\email{amowli@uci.edu}
\affiliation{%
  \institution{University of California, Irvine}
  \country{}
}

\begin{abstract}
Sparse attention is a core building block in many leading neural network models, from graph-structured learning to sparse sequence modeling. It can be decomposed into a sequence of three sparse matrix operations (3S): sampled dense-dense matrix multiplication (SDDMM), softmax normalization, and sparse matrix multiplication (SpMM). Efficiently executing the 3S computational pattern on modern GPUs remains challenging due to (a) the mismatch between unstructured sparsity and tensor cores optimized for dense operations, and (b) the high cost of data movement. 
Previous works have optimized these sparse operations individually or addressed one of these challenges. This paper introduces Fused3S, the first fused 3S algorithm that jointly maximizes tensor core utilization and minimizes data movement. Across real-world graph datasets, Fused3S achieves $1.6- 16.3\times$ and $1.5-14\times$ speedup over state-of-the-art on H100 and A30 GPUs. Furthermore, integrating Fused3S into Graph Transformer inference accelerates end-to-end performance by $1.05-5.36\times$, consistently outperforming all 3S baselines across diverse datasets (single and batched graphs) and GPU architectures. 
\end{abstract}



\maketitle

\section{Introduction}

Attention has become fundamental in machine learning models from transformers \cite{vaswani2017attention} to graph neural networks (GNNs) \cite{veličković2018graph, thekumparampil2018attention, GraphTransformer}.
However, its computational cost remains a bottleneck as we scale in sequence length and graph size.
While dense and block-sparse attention have benefited from hardware-aware algorithm design \cite{dao2022flashattention, dao2024flashattention2}, sparse attention--essential for graph-based learning and dynamic sparsity patterns--remains under-optimized on modern hardware accelerators. 
This inefficiency is especially pronounced on GPUs with tensor cores, which deliver peak throughput for dense matrix multiplications (or limited structured sparsity such as 2:4) with strict operand shapes.
In contrast, sparse operations involve irregular memory accesses and unstructured computation, making it poorly suited for current tensor core design.
As a result, tensor cores remain largely underutilized for sparse workloads.

Sparse attention can be decomposed into a sequence of three operations: Sampled Dense-Dense Matrix Multiplication (SDDMM) to compute attention scores, softmax normalization, and Sparse Matrix Multiplication (SpMM) to aggregate features. 
We refer to this computational pattern as \textbf{3S}, which recurs in GNNs \cite{GraphTransformer, veličković2018graph, thekumparampil2018attention}, sparse transformers \cite{lee2024sea, liu2022DSA}, and models that exploit dynamic sparsity. 

\setlength{\belowcaptionskip}{-1pt}
\begin{table*}[h!]
\centering
\caption{Summary of algorithms designed for 3S or its sub-operations (SDDMM or SpMM). }
\begin{tabular}{lccccccc}
\toprule
\textbf{Method}  & \textbf{Hardware} & \textbf{Format} & \textbf{Precision} & \multicolumn{2}{c}{\textbf{Kernels}} & \textbf{Fusion} & \textbf{3S} \\ 
\cline{5-6}
& & & & \textbf{SDDMM} & \textbf{SpMM} &  & \\ 
\hline
Sputnik \cite{Sputnik}  & CUDA & CSR & fp16, fp32 & \gdot & \gdot & \rdot & \gdot\\ 
RoDe \cite{RoDe} & CUDA & CSR & fp32, fp64 & \gdot & \gdot & \rdot  & \rdot \\ 
JigSaw\cite{Jigsaw}  & SPTC & Reorder-aware & fp16 & \rdot & \gdot & \rdot & \rdot\\ 
TCA-SpMM\cite{TCA-SpMM} & TC & CSR & fp16/fp32 & \rdot & \gdot & \rdot & \rdot\\ 
Magicube \cite{Magicube} & TC & SR-BCRS & int16, int8, int4 & \gdot & \gdot & \rdot & \gdot \\ 
SMaT\cite{SMaT} & TC & BCSR & fp16 & \rdot & \gdot & \rdot & \rdot\\ 
BSA-SpMM \cite{BSA-SpMM} & TC & CSR, Blocked-ELL & fp16 & \rdot & \gdot & \rdot & \rdot\\ 
Flash-LLM \cite{Flash-LLM} & TC & Tiled-CSL & fp16/fp32 & \rdot & \gdot & \rdot & \rdot \\ 
TC-GNN\cite{TCGNN} & TC & TCF & tf32 & \gdot & \gdot & \rdot & \rdot\\ 
DTC-SpMM\cite{DTC-SpMM} & TC & ME-TCF & tf32 & \rdot & \gdot & \rdot & \rdot \\ 
Acc-SpMM\cite{zhao2024accspmm} & TC & BitTCF & tf32 & \rdot & \gdot & \rdot & \rdot\\
FlashSparse\cite{shi2024flashsparse} & TC & ME-BCRS & fp16/tf32 & \gdot & \gdot & \rdot & \gdot\\
\hline
FusedMM\cite{FusedMM2021} & CPU & CSR & fp32, fp64 & \gdot & \gdot & \gdot & \rdot\\ 
DF-GNN\cite{liu2024dfgnn} & CUDA & CSR+COO, CSC & fp32 & \gdot & \gdot & \gdot & \gdot\\
\textbf{Fused3S (this paper)} & TC & BSB & fp16/fp32 & \gdot & \gdot & \gdot & \gdot\\ \bottomrule
\end{tabular}
\label{tab:algorithms}
\end{table*}

Prior efforts to accelerate the 3S pattern fall into two broad categories: (1) \emph{Individual kernel optimizations}, which improves the performance of one or more sparse operations (such as SDDMM and/or SpMM) in isolation using specialized sparse tensor formats and kernel-local optimizations \cite{TCGNN, DTC-SpMM, zhao2024accspmm, Flash-LLM, BSA-SpMM, SMaT, TCA-SpMM, Magicube, shi2024flashsparse}. These approaches incur unnecessary data movement when intermediate results are materialized in global memory.
(2) \emph{Kernel fusion}, which reduces memory traffic by combining the 3S operations into a single kernel. However, existing fused kernels for sparse attention are designed either for CPUs \cite{FusedMM2021} or CUDA cores \cite{liu2024dfgnn}, leaving tensor core acceleration untapped. 
As summarized in Table \ref{tab:algorithms}, no existing work fuses the 3S operations while targeting tensor cores. 

To bridge this gap, we propose \textbf{Fused3S}, the first fused sparse attention algorithm and kernel designed for GPU tensor cores. 
\kernel: (1) adopts a block-structured sparse format aligned with tensor core operand shapes, (2) fuses SDDMM, softmax, and SpMM into a single kernel to reuse intermediate results in registers and shared memory, and (3) implements a mixed precision pipeline with numerically stable online softmax to maximize performance while maintaining accuracy.

Our contribution can be summarized as follows.
\begin{itemize}
    \item \textbf{\kernel}\footnote{\url{https://github.com/HPCForge/Fused3S}}, an open-source kernel that simultaneously exploits kernel fusion \emph{and} tensor core utilization for the 3S sparse computational pattern. 
    \item The fused algorithm is designed to be fully on-chip with high parallelism. This is achieved using multi-level tiling with efficient block- and warp-level work partitioning to avoid global-memory synchronization. Reordering and register-level remapping optimizations improve load balance and memory accesses for irregular graphs.
    \item Across real-world graph datasets, \kernel\ achieves $1.6-16.3\times$ and $1.5-14\times$ speedups over DF-GNN \cite{liu2024dfgnn}, FlashSparse \cite{shi2024flashsparse}, and PyG \cite{FeyLenssen2019PyG} on H100 and A30 GPUs respectively. 
    \item Integrated into the Graph Transformer \cite{GraphTransformer} implemented in DGL \cite{wang2019deep}, \kernel\ achieves $1.05-5.36\times$ end-to-end inference speedup over state-of-the-art 3S baselines across graph datasets and GPUs. 
\end{itemize}

\section{Background}
\subsection{Computational Pattern in Sparse Attention}
A common computation in machine learning models from graph-structured learning to sparse sequence modeling is the \textbf{3S} pattern: a sequence of SDDMM, softmax normalization, and SpMM. 
3S can be formulated as:
\begin{equation}
\M{O} = \text{softmax}(\M{Q}\M{K}^T \odot \M{A})\M{V}
\label{eqn:3s}
\end{equation}

where $\M{Q}$, $\M{K}$, $\M{V}$, and $\M{O} \in \mathbb{R}^{N\times d}$ are dense matrices and $\M{A} \in \mathbb{R}^{N\times N}$ is a sparse matrix that defines attention patterns (\eg adjacency or masking). 
Equation \ref{eqn:3s} can be decomposed into three operations: 

\begin{enumerate}
    \item \textbf{SDDMM:} Compute attention scores $\M{S} = \M{Q}\M{K}^T \odot \M{A}$, where the dense-dense multiplication $\M{Q}\M{K}^T$ is computed only for non-zeros in $\M{A}$.
    \item \textbf{Softmax:} Normalize the scores row-wise $\M{E} = \text{softmax}(\M{S})$.
    \item \textbf{SpMM:} Aggregate output $\M{O} = \M{E}\M{V}$.
\end{enumerate}

This 3S pattern appears in several popular architectures.

\textbf{Graph Attention Network (GAT).}
In GATs \cite{veličković2018graph}, nodes in a graph attend selectively to their neighbors using $\M{A}$ as the adjacency matrix. A typical formulation of GAT attention is: 
\begin{equation}
\M{O} = \text{softmax}\Big(\text{LeakyReLU}([\M{W}\M{H} || \M{W}\M{H}]) \odot \M{A}\Big)(\M{W}\M{H}),
\end{equation}

where $\M{H}$ is the input node features, $\M{W}$ is a learnable weight matrix, and $||$ denotes concatenation. (1) SDDMM computes the unnormalized attention coefficients between nodes. (2) Softmax normalizes the attention coefficients across all neighbors of a node. (3) SpMM aggregates the transformed features of neighboring nodes, weighted by the normalized attention coefficients.

\textbf{Attention-based Graph Neural Network (AGNN).}
AGNN \cite{thekumparampil2018attention} introduces a dynamic, adaptive attention. We can  formulate AGNN as:
\begin{equation}
\M{O} = \text{softmax}\Big(\beta^{(t)} \cos(\M{H}^{(t)}, \M{H}^{(t)^T}) \odot \M{A}\Big)\M{H}^{(t)}
\end{equation}
where $\M{A}$ includes self-loops, $\beta^{(t)}$ is a learnable parameter for layer $t$, and $\cos(\cdot,\cdot)$ denotes the cosine similarity. Here, $\M{Q} = \M{K} = \M{V} = \M{H}^{(t)}$, with the SDDMM step computing scaled cosine similarities.

\textbf{Graph Transformer (GT).}
GTs \cite{GraphTransformer, shirzad2023exphormer, zhang2024torchgt} extend attention to entire graphs, treating nodes as tokens. A representative formulation is:
\begin{equation}
\M{O} = \text{softmax}\Big((\M{W}_Q\M{H})(\M{W}_K\M{H})^T \odot \M{A}\Big)(\M{W}_V\M{H}),
\end{equation}
where $\M{W}_Q$, $\M{W}_K$, and $\M{W}_V$ are learnable projections for queries, keys, and values. Unlike standard transformers, GTs explicitly encode graph structure through $\M{A}$.

Recent surveys \cite{ller2024attending, shehzad2024graphtransformerssurvey} note that many GT models use dense global attention, augmented with structural bias (\eg node degrees, shortest paths) to avoid over-smoothing and improve expressivity. However, this approach can be computationally expensive for large graphs, motivating the need for sparse attention. 

\textbf{Sparse Transformers.}
Sparse transformers \cite{SparseTransformers} reduce the quadratic complexity of attention by applying a sparse mask $\M{M}$.
\begin{equation}
\M{O} = \text{softmax}\Big((\M{W}_Q\M{X})(\M{W}_K\M{X})^T \odot \M{M}\Big)(\M{W}_V\M{X}),
\end{equation}
where $\M{X}$ is the input sequence. 
The mask $\M{M}$ determines token interactions and may be static or dynamically generated. Static masks \cite{BigBird, beltagy2020longformer} impose structured sparsity (\eg block-diagonal and block-sparse) that is GPU friendly. Dynamic variants \cite{lee2024sea, liu2022DSA} compute $\M{M}$ on-the-fly enabling context-aware sparsity. While dynamic masks often improve accuracy, they introduce irregular sparsity that is difficult to optimize efficiently. 

Although GATs use fixed $\M{A}$ and sparse transformers generate $\M{M}$ dynamically, these diverse models share the same 3S bottleneck: computing and applying sparse attention on modern hardware accelerators. Our work targets this unifying 3S abstraction to develop a hardware-optimal algorithm.

\subsection{Tensor Core and Operand Shapes}
Tensor Cores (TCs) are specialized hardware units on NVIDIA GPUs designed for high-throughput matrix multiply and accumulate operations. 
Since their introduction in 2017, FP16 throughput using TCs has increased from 125 TFLOPS on V100 to 990 TFLOPS on H100, an improvement of $8\times$ in 5 years \cite{V100DataSheet, GH200DataSheet}. 
This rapid progression has significantly improved the performance of dense matrix computations.

There are two primary programming interfaces for TCs: CUDA \texttt{wmma} (Warp Matrix Multiply Accumulate) and PTX \texttt{mma} (Matrix Multiply Accumulate) instructions. The choice of interface depends on specific optimization goals. The PTX \texttt{mma} is a lower-level interface that allows direct operand loading from global memory (HBM) into registers, bypassing shared memory. This provides finer-grained control and can be advantageous for workloads with limited data reuse. 
In contrast, \texttt{wmma} operates at a higher abstraction level, requiring both input matrices to be explicitly staged in shared memory before loading into registers.

\begin{table}[h]
\centering
\caption{Precision formats and operand shapes on Tensor Cores. Here $m, n, k$ denote tile dimensions for matrix multiplication.
}
\begin{tabular}{|c|l|l|}
\hline
\textbf{Precision} & \textbf{Type} & \textbf{Operand Shapes} \\ \hline
\multirow{3}{*}{\texttt{FP16}} & \texttt{wmma} & m16n16k16, m8n32k16, m32n8k16 \\ \cline{2-3} 
                              & \texttt{mma}  & m8n8k4, m16n8k8, \textbf{m16n8k16} \\ \hline
\multirow{2}{*}{\texttt{BF16}} & \texttt{wmma} & m16n16k16, m8n32k16, m32n8k16 \\ \cline{2-3} 
                              & \texttt{mma}  & m16n8k8, \textbf{m16n8k16} \\ \hline
\multirow{2}{*}{\texttt{TF32}} & \texttt{wmma} & m16n16k8 \\ \cline{2-3} 
                              & \texttt{mma}  & m16n8k4, m16n8k8 \\ \hline
\multirow{2}{*}{\texttt{FP8}}  & \texttt{wmma} & -- \\ \cline{2-3} 
                              & \texttt{mma}  & m16n8k32, \textbf{m16n8k16} \\ \hline
\end{tabular}
\label{table:wmma_mma_precision}
\end{table}

TCs support various operand shapes and precision formats summarized in Table \ref{table:wmma_mma_precision}.
These shapes dictate how input matrices are partitioned into tiles and strongly influence performance.
For sparse computations, the optimal tile size may not always align with the hardware's peak capability. 
The architectural trend towards "dense TCs" suggests larger tile shapes to maximize TC utilization but this can lead to more zero computations when applied naively to sparse matrices. 
Smaller tile shapes reduce the occurrence of zeros and improve compute density on sparse data, but may result in underutilized TCs.
Among the available configurations, the \texttt{m16n8k16} tile shape emerges as a practical compromise, supported in multiple precision formats (FP16, BF16, FP8). 
\section{\kernel}
This section presents the design of \kernel. We first describe how the sparse matrix $\M{A}$ is stored in a block-structured format tailored to tensor core operand shapes. Then, we detail the \kernel\ kernel resulting in Algorithm \ref{alg:fused3s} and highlight the key optimizations: 

\begin{itemize}
    \item Fusing the 3S operations (SDDMM, Softmax, SpMM) into a single kernel using multi-level tiling/blocking to reduce memory traffic and enable on-chip data reuse.
    \item Incremental softmax computation to support large attention matrices.
    \item Warp-level parallelism for SIMT-friendly execution.
    \item Permuted data layouts and register-level remapping to enable coalesced memory access patterns. 
\end{itemize}

\subsection{Sparse Format for Tensor Cores}
\label{subsec:bsb}

We introduce the \textbf{Binary Sparse Block (BSB)} format to efficiently map a sparse matrix $A$ onto tensor cores. BSB extends the Memory-Efficient Tensor Core Format (ME-TCF) \cite{DTC-SpMM}, which in itself builds on the TC-GNN Compressed Format (TCF) \cite{TCGNN}. Like block-CSR (BCSR), these formats use a block layout with local indexing but are specifically designed for tensor core operand shapes. 

\begin{figure}[h]
    \centering
    \includegraphics[width=1\linewidth]{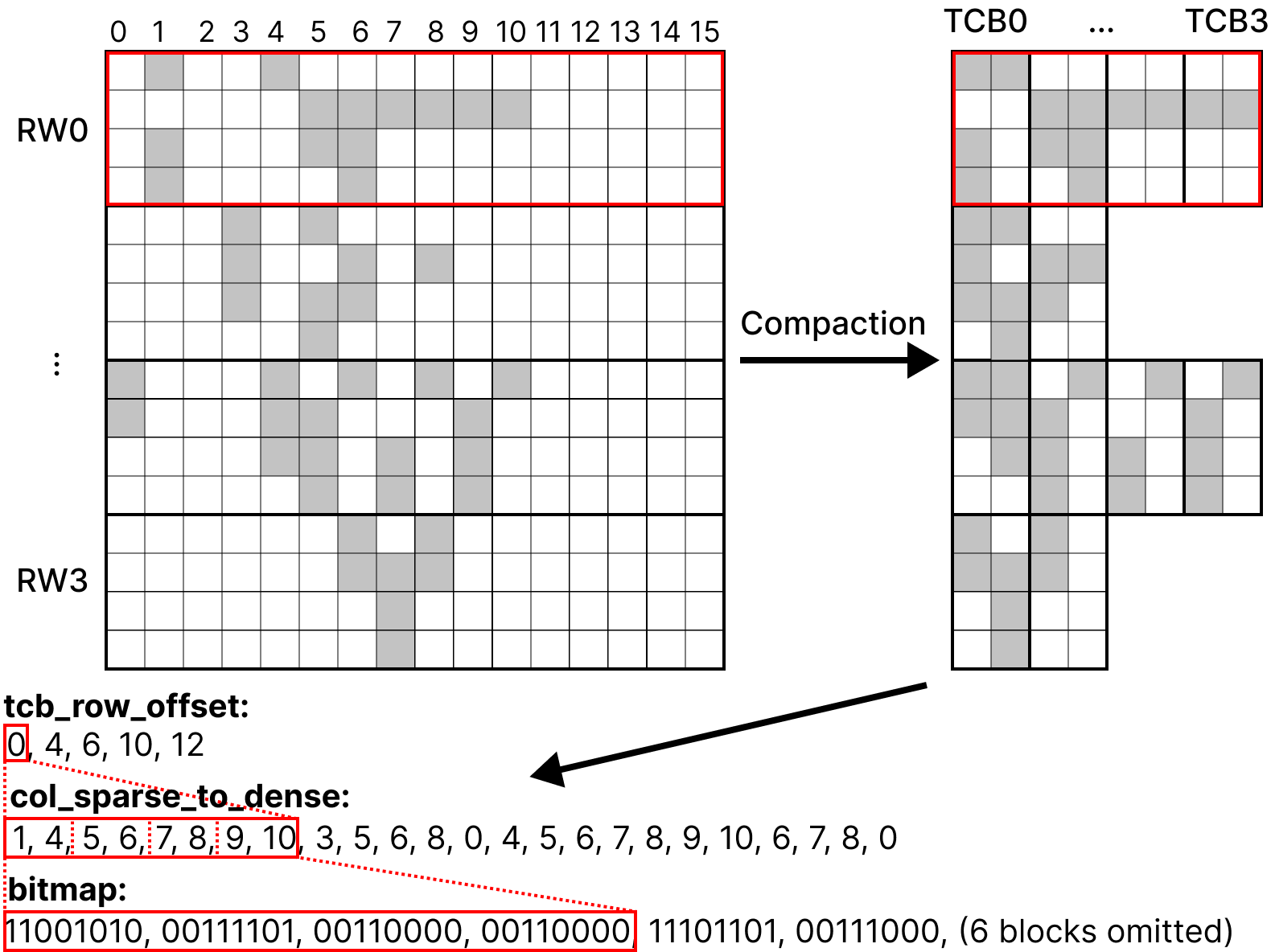}
    \caption{\text{Binary Sparse Block (BSB)} format. The TCB size in this example is $4\times2$ while in practice the size is larger (\ie $16\times8$). Red boxes highlight how the first row window in compacted, tiled, and stored in BSB format.}
    \label{fig:BSB}
\end{figure}

The construction of the BSB format proceeds as follows and is illustrated in \Cref{fig:BSB}:
\begin{enumerate}
    \item Divide the sparse matrix into \emph{row windows} (RW) of size $r$. 
    \item Within each RW, eliminate columns containing only zeros to increase compute density.
    \item Partition the compacted RW into \emph{tensor core blocks} (TCB) of shape $r \times c$, where $r$ and $c$ match supported \texttt{mma} tile sizes (\eg $16 \times 8$ in Table \ref{table:wmma_mma_precision}). 
    \item We maintain three data structures: 
    \begin{itemize}
        \item \texttt{tcb\_row\_offset (tro)}: Number of TCBs per RW. 
        \item \texttt{col\_sparse\_to\_dense (sptd)}: Mapping from original to compacted column indices per RW.
        \item \texttt{bitmap}: A fixed-size bitmask encoding the sparsity pattern in each TCB. 
    \end{itemize}
\end{enumerate}

ME-TCF uses two arrays to store non-zero elements: one for the number of non-zero elements in each TCB and another to store the local index of each nonzero element in all TCBs. 
We make the observation that adjacency matrices and attention masks exhibit binary-valued sparsity.
Unlike ME-TCF and TCF, which represent the location of nonzeros using integer indices, BSB encodes the $r \times c$ TCB using a single binary bitmap.
For example, a $16\times8$ TCB requires only 128 bits to represent its sparsity pattern, eliminating indexing overhead. 

Table \ref{tab:format_compare} compares the memory footprint of various sparse formats. The formats differ in how they organize sparsity (row-based vs. block-based) and whether they store explicit nonzero values.
Row-based formats (such as CSR) are incompatible with tensor cores due to irregular access patterns. 
General-purpose block formats such as BCSR \cite{im2004sparsity} and its variants \cite{shi2024flashsparse,Magicube, SMaT} improve locality but explicitly store both nonzero values and their positions.
In contrast, formats such as TCF \cite{TCGNN}, ME-TCF \cite{DTC-SpMM}, BitTCF \cite{zhao2024accspmm}, and our BSB are designed for tensor cores. These formats align blocks with MMA tile shapes and assume binary sparsity.
BSB further reduces overhead by encoding block sparsity with a fixed-size bitmap. Unlike BCSR, BSB compacts columns within row windows to increase density and reduce the total number of blocks.

\begin{table}[h]
\centering
\caption{Comparison of sparse formats. \texttt{row}: row-based, \texttt{blk}: generic block-based, \texttt{mma}: MMA-tile-aligned. Matrix size is $N \times N$ with $z$ nonzeros. Row window height is $r$; $b$: number of blocks, $bc$: stored columns after compaction (if any), and $rc$: elements per block. Sizes assume 32-bit indices and values unless format is binary.}
\label{tab:format_compare}
\begin{tabular}{l|c|l|l}
\textbf{Format} & \textbf{Type} & \textbf{Memory Footprint} & \textbf{NZ Value} \\\hline
CSR        & row & $32(N + 2z)$ & fp32 \\\hline
SR-BCSR    & blk & $32(\frac{2N}{r} + bc + brc)$ & fp32 \\
ME-BCRS    & blk & $32(\frac{N}{r} + bc + brc)$ & fp32 \\
BCSR       & blk & $32(\frac{N}{r} + b + brc)$ & fp32 \\\hline
TCF        & mma & $32(\frac{N}{r} + N + 3z)$ & binary \\
ME-TCF     & mma & $32(\frac{N}{r} + b + z) + 8z$ & binary \\
BitTCF     & mma & $32(\frac{N}{r} + b + z) + z$ & binary \\
BSB (ours) & mma & $32(\frac{N}{r} + bc) + brc$ & binary \\
\end{tabular}
\end{table}

\subsection{Fusion and Thread-block Parallelization}

\begin{table}[h]
\centering
\caption{Notation used in Algorithm \ref{alg:fused3s}.}
\label{table:notations}
\begin{tabular}{|l|l|}
\hline
\textbf{Symbol} & \textbf{Definition} \\
\hline
$\mathbf{Q}, \mathbf{K}, \mathbf{V} \in \mathbb{R}^{N\times d}$   & Query, key, value matrices \\
$\mathbf{A} \in \mathbb{R}^{N\times N}$   & Sparse matrix (\eg adjacency or mask) \\
$\mathbf{S} \in \mathbb{R}^{N\times N}$   & Attention score matrix \\
$\mathbf{E} \in \mathbb{R}^{N\times N}$   & Row-wise normalized score matrix \\
$\mathbf{O} \in \mathbb{R}^{N\times d}$   & Output matrix \\
$\mathbf{Q}_{i} \in \mathbb{R}^{r\times d}$ & Query block \\
$\hat{\M{K}}, \hat{\M{V}} \in \mathbb{R}^{t c \times d}$ & Gathered rows of key and value matrices \\
$\mathbf{S}_{i} \in \mathbb{R}^{r \times c W}$ & Attention score block\\
$\mathbf{E}_{i} \in \mathbb{R}^{r \times c W}$ & Normalized score block \\
$\mathbf{O}_{i} \in \mathbb{R}^{r\times d}$ & Output block \\
$\mathbf{m}_{o} \in \mathbb{R}^{r}$       & Row-wise max scores \\
$\mathbf{l}_{o} \in \mathbb{R}^{r}$       & Row-wise softmax normalization factor \\
\hline
$r,c$                & Dimensions of tensor core block \\
$t$                  & Number of TCBs in row window $i$ \\
$W$                  & Number of warps per thread block\\
$N$                  & Number of rows/nodes\\
$d$                  & Feature dimension \\
\hline
\end{tabular}
\end{table}

\begin{algorithm}[h!]
    \small
    \caption{\textsc{\kernel}} 
    \label{alg:fused3s}
    \begin{algorithmic}[1]
    \Require{$\M{A}$ in BSB format: \texttt{tro}, \texttt{sptd}, \texttt{bitmap}; $\M{Q}, \M{K}, \M{V} \in \mathbb{R}^{N\times d}$}
    \Ensure{$\M{O} = \text{softmax}(\M{Q}\M{K}^{\Tra}\odot\M{A}) \M{V} \in \mathbb{R}^{N\times d}$}
        \State Divide $\M{Q}$ into $T_r=\lceil{\frac{N}{r}}\rceil$ blocks \{$\M{Q}_1$ ... $\M{Q}_{T_r}$\}, each of size $r\times d$ 
        \State Divide $\M{O}$ into $T_r$ blocks \{$\M{O}_1$ ... $\M{O}_{T_r}$\}, each of size $r\times d$
        \For{$i = 1 \; \text{to} \; T_r$} 
            \State Initialize $\V{m}_{o} = -\infty,\, \V{l}_{o} = 0 \in \mathbb{R}^{r}$, ${\M{O}_{i}} = 0 \in \mathbb{R}^{r\times d}$ in \texttt{fp32}
            \State Load $\M{Q}_i$ from HBM to SMEM 
            \State $t = \texttt{tro}[i+1] - \texttt{tro}[i]$ 
            \State \V{c} = getColumnVectorIndex(\texttt{sptd}, $i$)
            \State $\hat{\M{K}}, \hat{\M{V}} \in \mathbb{R}^{tc\times d}$ = select rows of $\M{K}, \M{V}$ according to \V{c} 
            \State Divide $\hat{\M{K}}$ into $T_c = \lceil\frac{t}{W}\rceil$ blocks $\{\hat{\M{K}}_1 ... \hat{\M{K}}_{T_c}\}$, each of size $Wc \times d$
            \State Divide $\hat{\M{V}}$ into $T_c$ blocks $\{\hat{\M{V}}_1 ... \hat{\M{V}}_{T_c}\}$, each of size $Wc \times d$
            \For{$j = 1 \; \text{to} \; T_c$} 
                \State{// SDDMM}
                \State $\M{S}_i$ = TBGemm($\M{Q}_i,\,\hat{\M{K}}_j^\Tra$,\,\M{0})
                \State Apply \texttt{bitmap} mask to $\M{S}_i$
                \State{// Online Softmax}
                \State $\V{m}_i = \max{(\V{m}_o, \texttt{rowmax}(\M{S}_i))}$
                \State $\M{E}_i = e^{\M{S}_i - \V{m}_i}$
                \State $\V{l}_o = \text{diag}(e^{\V{m}_o - \V{m}_i})\V{l}_o + \texttt{rowsum}(\M{E}_i)$
                \State Store $\M{E}_i$ (cast to fp16) in SMEM
                \State{// SpMM}
                \State $\M{O}_i = \text{diag}(e^{\V{m}_o - \V{m}_i}) \M{O}_i$
                \State $\M{O}_i$ = TBGemm($\M{E}_i,\,\hat{\M{V}}_j,\,\M{O}_i$)
                \State $\V{m}_o = \V{m}_i$ 
            \EndFor
            \State Write $\M{O}_i = \text{diag}(\V{l}_o)^{-1}\M{O}_i$ to HBM
        \EndFor
    \end{algorithmic}
\end{algorithm}

\begin{algorithm}[h!]
    \small
    \caption{\textsc{TBGemm}} 
    \label{alg:TBGemm}
    \begin{algorithmic}[1]
    \Require {MMA tile: $(m, n, k)$; $\M{A} \in \mathbb{R}^{m \times K}$ in SMEM (fp16),  $\M{B} \in \mathbb{R}^{K \times P}$ in HBM (fp16), $\M{D} \in \mathbb{R}^{m \times P}$ in SMEM (fp32)}
    \Ensure{$\M{C} = \M{A} \M{B} + \M{D} \in \mathbb{R}^{m \times P}$ in fp32}
        \State Divide $\M{A}$ into $T_g =\lceil \frac{K}{k}\rceil$ tiles $\{\M{A}_1, ... , \M{A}_{T_g}\}$, each of size $m \times k$
        \State Divide $\M{B}$ into $T_h = \lceil \frac{P}{n}\rceil$ tiles $\{\M{B}_1, ... , \M{B}_{T_h}\}$, each of size $K \times n$
        \State Divide $\M{D}$ into $T_h$ tiles $\{\M{D}_1, ... , \M{D}_{T_h}\}$, each of size $m \times n$
        \For{$i = 1$ to $T_h$}
            \State Load $\M{D}_i$ from SMEM to registers as $\M{C}_i$
            \State Divide $\M{B}_i$ into $T_g$ tiles $\{\M{B}_{i1}, ... , \M{B}_{iT_g}\}$, each of size $k \times n$
            \For{$j = 1$ to $T_g$}
                \State Load $\M{A}_j$ from SMEM to registers
                \State Load $\M{B}_{ij}$ from HBM to registers
                \State $\M{C}_i = \texttt{mma}(\M{A}_j, \M{B}_{ij}, \M{C}_i)$ // Tensor Core operation 
            \EndFor
        \EndFor
        \State \Return $\M{C}$
    \end{algorithmic}
\end{algorithm}



\Cref{alg:fused3s} describes the \kernel kernel. 
The input and output matrices $\M{Q}$ and $\M{O}$ are divided into row blocks (line 1-2), each assigned to a thread block.
Each thread block loads its corresponding $\M{Q}_i$ into shared memory (line 5), which is reused across TCBs in the row window. 
The number of TCBs in the $i$-th row window is determined using $\texttt{tro}$ (line 6).
Thread blocks then extract the column indices \V{c} that define the nonzero pattern in $\M{A}_i$ using the \texttt{sptd} map (line 7).
These indices are used to gather rows from $\M{K}$ and $\M{V}$ (line 8), which are then partitioned into warp-aligned blocks (lines 9-10). 
Unlike $\M{Q}_i$, which is reused across all warps, $\hat{\M{K}}$ and $\hat{\M{V}}$ are only accessed once per row window and loaded directly from HBM into registers without staging in shared memory.

The inner loop (lines 11-23) fuses the 3S operations.
SDDMM is executed using a warp-level \texttt{TBGemm} (line 13), computing attention scores $\M{S}_i$, which are masked with the sparse bitmap from the BSB format (line 14).
Softmax is computed incrementally using a numerically stable online variant adapted from FlashAttention-2~\cite{dao2024flashattention2} (lines 16–18).
We track the running row-wise max $\V{m}_o$ and normalization factor $\V{l}_o$, and apply exponential rescaling across blocks to preserve numerical stability and ensure correctness, despite the blockwise computation. All scaling is done in fp32.
Normalized scores $\M{E}_i$ are cast to fp16 and stored in shared memory (line 19).
SpMM proceeds by rescaling the accumulated output block $\M{O}_i$ and invoking a second \texttt{TBGemm} (lines 21–22).
After processing all blocks in the row window, the final output block $\M{O}_i$ is normalized and written to HBM (line 24).

The \texttt{TBGemm} kernel in Algorithm \ref{alg:TBGemm} is a core primitive used in SDDMM and SpMM (lines 13 and 22). It partitions input blocks into tensor core compatible tiles, loads operands into registers, and issues MMA instructions to perform high-throughput matrix multiply-accumulate.

\begin{figure}[h]
    \centering
    \includegraphics[width=1\linewidth]{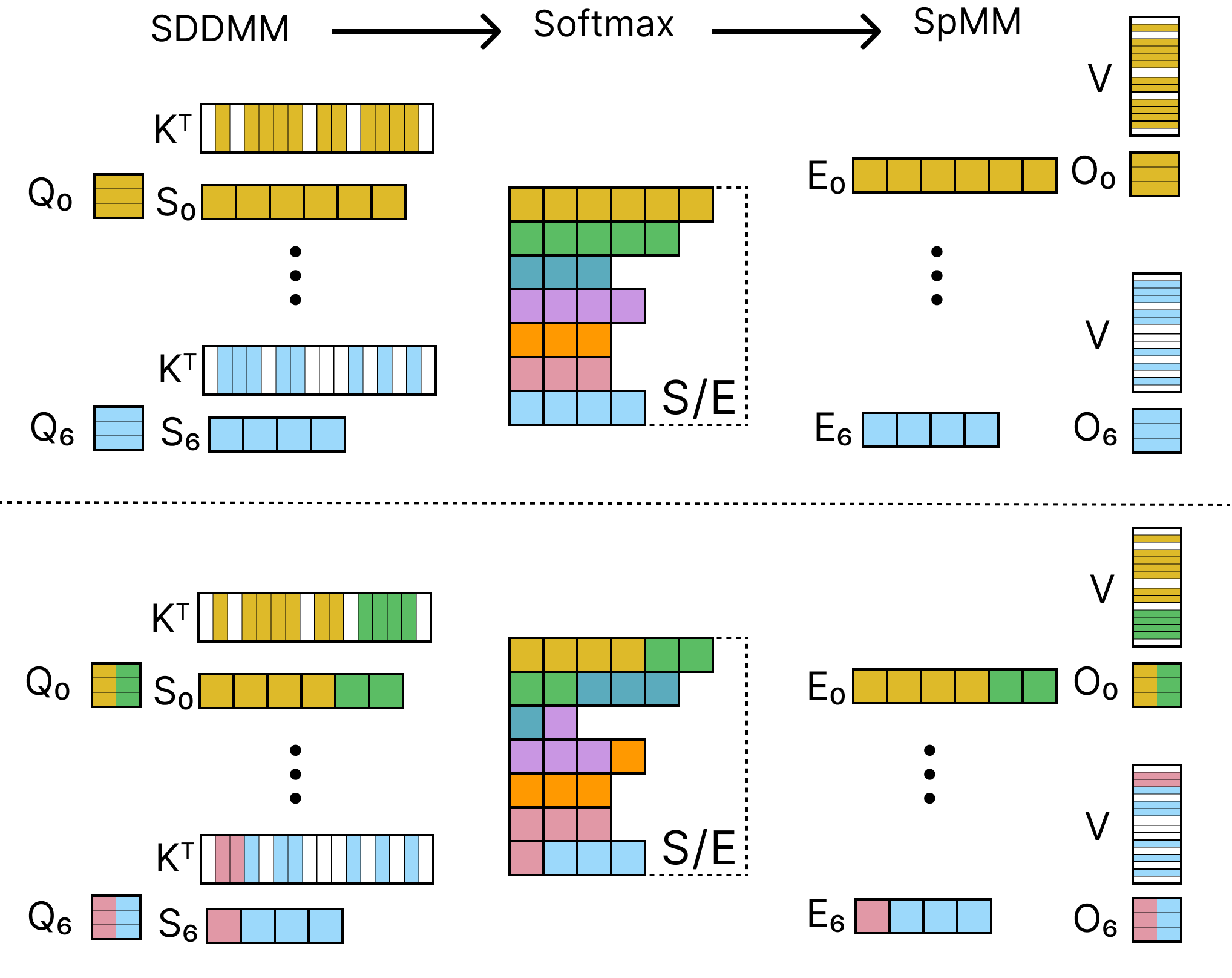}
    \caption{Comparison of node-parallel (top) and edge-parallel (bottom) strategies. Different colored blocks are owned by different thread blocks. In edge-parallel, blocks of $\M{Q}$ or $\M{O}$ are shaded with multiple colors if shared by multiple thread blocks. The figure is divided vertically into three stages: (1) SDDMM, (2) data distribution of $\M{S}$ and $\M{E}$ for softmax, and (3) SpMM. }
    \label{fig:nodeVsEdgeParallel}
\end{figure}

There are two ways to parallelize the 3S computation: \textbf{node-parallel} and \textbf{edge-parallel}. The distinction lies in how $\M{S}$, the output of SDDMM, is distributed among thread blocks, as illustrated in \Cref{fig:nodeVsEdgeParallel}. Although these terms are coined in the context of graph attention, the concepts apply more broadly to sparse attention.

\textbf{Node-Parallel Fusion.}
In node-parallel, each thread block is assigned a fixed set of rows in $\M{S}$ (\ie a subset of nodes in the graph). 
As seen in \Cref{alg:fused3s}, softmax requires row-wise reductions (max and sum), which can be computed locally within each thread block.
This enables the subsequent SpMM to also be executed within the same thread block.
The primary advantage of node-parallel is independence: each thread block owns all data needed for its rows of softmax and SpMM, avoiding inter-block synchronization.
This is depicted at the top of \Cref{fig:nodeVsEdgeParallel}, where thread blocks operate on disjoint rows.

\textbf{Edge-Parallel Fusion.}
Edge-parallel distributes computation across thread blocks based on TCBs (\ie edge-level granularity). 
This achieves better load balancing, especially for datasets where the number of TCBs per RW (i.e., node degree in graphs) varies widely (see \Cref{tab:tcb_deciles}). 
Such variance is common in real-world graphs due to their power-law degree distribution. 
By evenly distributing TCBs among thread blocks, edge-parallel ensures uniform workload across SMs. 
However, it introduces significant synchronization overhead. 
Since rows of $\M{S}$ (and hence $\M{O}$) may be computed by multiple thread blocks, softmax and SpMM must coordinate through global synchronization or atomic updates to HBM--both of which incur performance penalties. 
As shown in the bottom of \Cref{fig:nodeVsEdgeParallel}, rows may be fragmented across multiple thread blocks.
Prior work \cite{DTC-SpMM} reports that edge-parallel SpMM is 20\% slower than node-parallel on average. 
Since \kernel fuses softmax and SpMM, it requires an additional global synchronization for softmax, making edge-parallel even less attractive.

\textbf{Load Balancing via Row Window Reordering.}
\kernel fuses the 3S operations into a single kernel to reduce memory traffic, so minimizing global synchronization is important for performance. 
For this reason, we adopt node-parallel fusion. 

By default, we assign each \rowWindow\ to one thread block. In \Cref{alg:fused3s}, this corresponds to the outer loop (line 3), with each iteration executed in parallel by thread blocks. This can lead to load imbalance across thread blocks. We visualize the performance impact in \Cref{fig:SM_Active_Comp}, which shows that some SMs remain active long after the others have finished execution.

To mitigate this, we perform \emph{row window reordering}, where \rowWindows are sorted in decreasing order of \tcBlock count.
This prioritizes denser \rowWindows earlier in the kernel execution when more \rowWindows are available to be assigned to other thread blocks. Lightweight \rowWindows\ that complete quickly are deferred to the end. 
This scheduling policy improves SM utilization and reduces kernel tail latency.
Importantly, this reordering is performed during preprocessing, alongside sparse matrix compaction, and adds negligible overhead per input graph.
By combining node-parallel fusion with sparse layout optimizations, \kernel\ maximizes memory efficiency while maintaining scalability on irregular graphs.

\subsection{Warp Partitioning Strategies}\label{sec:warpPartitioning}

\begin{figure}[h!]
    \centering
    \includegraphics[width=1\linewidth]{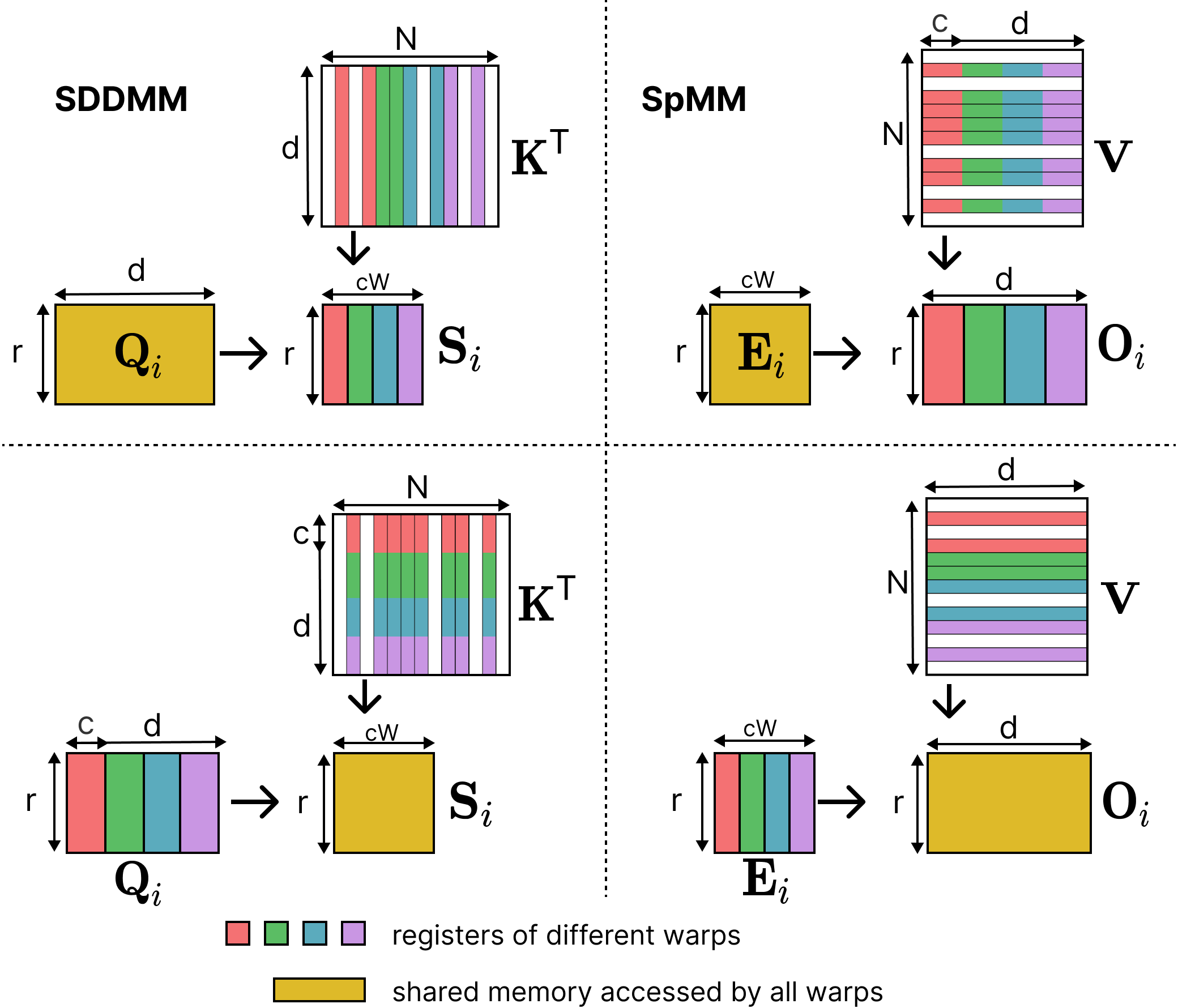}
    \caption{Work partitioning among warps within a thread block.
    Top: split-column (column blocks of $\M{K}^T$ and $\M{V}$ are divided among warps). Each warp independently computes a $r\times c$ tile of $\M{S}$ and $\M{O}$. Bottom: split-row (row blocks of $\M{K}^T$ and $\M{V}$ are divided among warps). All warps collaborate to compute each $r\times c$ tile of $\M{S}$ and $\M{O}$.}
    \label{fig:warpParallel}
\end{figure}

We explore two strategies to partition work among warps within a thread block: \textbf{split-column} and \textbf{split-row}. \Cref{fig:warpParallel} illustrates these two approaches for SDDMM and SpMM, highlighting each warp's data access pattern and its use of shared memory and registers. 

In the split-column scheme (top), the columns of the right-hand-side matrix ($\hat{\M{K}}^\Tra$ in SDDMM and $\hat{\M{V}}$ in SpMM) are divided among warps. Each warp computes a distinct $r \times c$ tile of the intermediate matrix $\M{S}$ and output matrix $\M{O}$. 
The advantage of this scheme is that warps operate on independent tiles, eliminating the need for inter-warp synchronization. 
However, each warp must access the entire $\M{Q}_i$ row block (in SDDMM) or $\M{E}_i$ (in SpMM), increasing memory pressure. 
The number of active warps in split-column is bounded by the number of \tcBlocks per \rowWindow ($t$ in line 9 of \Cref{alg:fused3s}). When $t$ is small, there may be insufficient warp-level parallelism to hide the latency of memory accesses. 

In the split-row scheme (bottom), the rows of $\hat{\M{K}}^\Tra$ and $\hat{\M{V}}$ are partitioned among warps. All warps within a thread block cooperate to compute each $r \times c$ tile of $\M{S}$ and $\M{O}$. 
This approach reduces the memory footprint per warp---each warp only loads a fragment of $\M{Q}_i$ or $\M{E}_i$.  
However, this comes at the cost of warp synchronization or atomic operations to aggregate the partial results in shared memory. 
In addition, the number of warps in split-row is constrained by the feature dimension $d$. For small $d$, limited parallelism may reduce the ability to hide the latency of irregular memory accesses of $\M{K}$ and $\M{V}$.

The trade-offs between these schemes involve memory access patterns, synchronization cost, register/shared memory pressure, and degree of parallelism exposed at the warp level. \Cref{alg:fused3s} is based on node-parallel thread block partitioning (\ie $T_r$ is partitioned among thread blocks) with split-column warp partitioning (\ie $T_c$ is partitioned among warps). 
We chose split-column as the default because, for typical attention workloads, the cost of accessing the entire $r \times d$ row block is often less significant than the cost of inter-warp synchronization. Furthermore, split-column maps naturally to the SIMT execution model of GPUs, enabling efficient parallel computation on independent tiles. 

\subsection{Data Layout and Memory Accesses}\label{sec:memAccess}
Efficient memory access is important for high performance, especially for sparse operations, which are often memory-bandwidth bound. We analyze the memory access patterns in SDDMM and SpMM at the thread block and warp levels, and describe a permuted data layout to improve memory coalescing. 

\begin{figure*}[h]
    \centering
    \includegraphics[width=\linewidth]{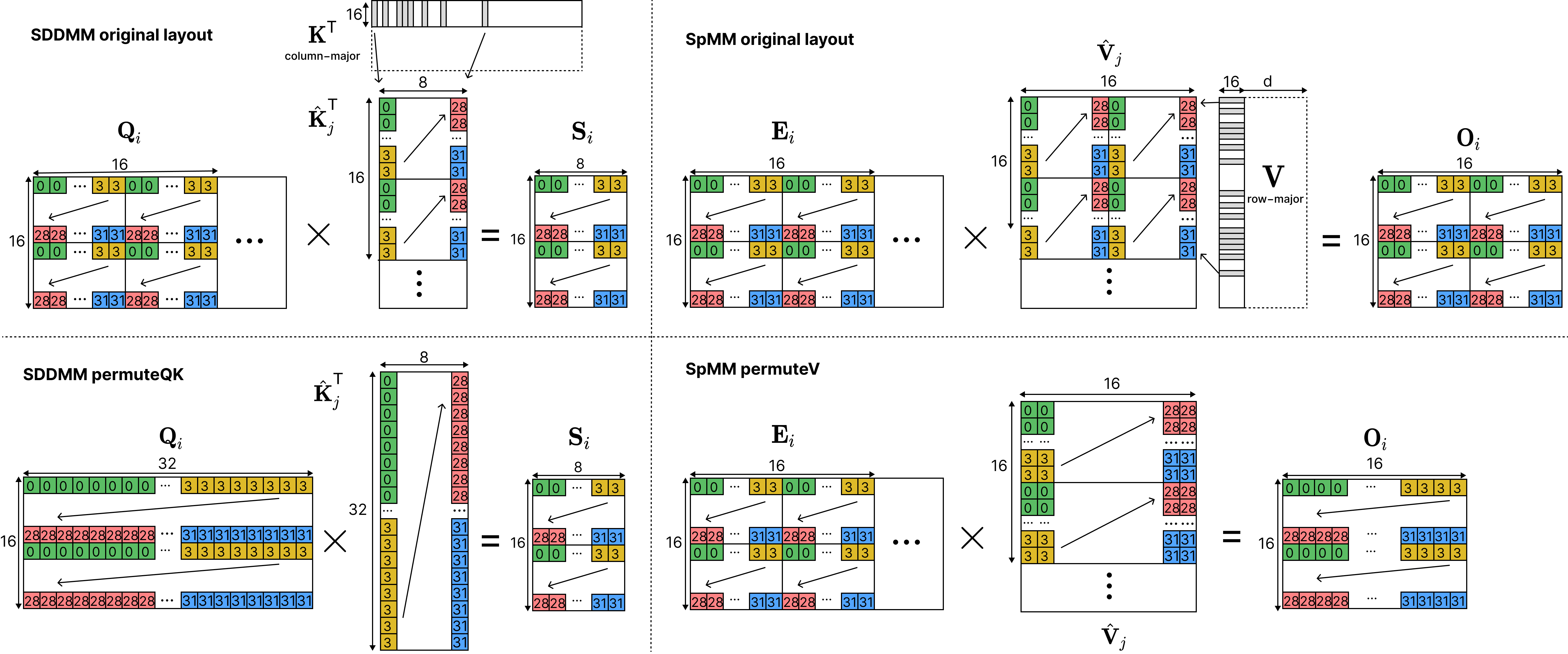}
    \caption{Register remapping in SDDMM (left) and SpMM (right). Top: original layouts. Bottom: permuted layouts.}
    \label{fig:regRemap}
\end{figure*}

\textbf{SDDMM.} 
Within each thread block, all warps access the same row block of $\M{Q}$ (see \Cref{fig:warpParallel}). Given this data reuse, $\M{Q}_i$ is copied from HBM to shared memory once per thread block (line 5 of \Cref{alg:fused3s}). 
In contrast, $\hat{\M{K}}_j^\Tra$ is partitioned column-wise among warps and is not reused within the thread block. Therefore, it is loaded directly from HBM into registers.
Since $\hat{\M{K}}$ is formed by gathering non-contiguous rows from $\M{K}$ based on the column indices of nonzeros in $\M{A}_i$, this leads to uncoalesced memory accesses.
Furthermore, the PTX \texttt{mma} interface requires tensor core operands to follow specific alignment and layout constraints.
This results in each thread issuing multiple 32-bit load instructions from scattered addresses as seen at the top-left of \Cref{fig:regRemap}. 

To address this, we apply a \emph{register remapping} optimization. As illustrated in the bottom-left of \Cref{fig:regRemap}, we optimize the memory access such that each thread issues a single 128-bit wide load.
This is equivalent to permuting the columns of $\hat{\M{K}}_j^\Tra$. 
To preserve correctness of the output and to be compatible with SpMM, we apply the same permutation to the columns of $\M{Q}_i$.
This maintains the mathematical operation while optimizing memory access.
Softmax does not incur additional data movement, as the result $\M{S}_i$ of SDDMM is already resident in registers. 
After softmax, each warp writes its slice of $\M{E}_i$ to shared memory, where it is reused by SpMM.

\textbf{SpMM.}
In SpMM, $\M{E}_i$ is already in shared memory. 
The matrix $\hat{\M{V}}$ is gathered from rows of $\M{V}$ using the same indices as for $\hat{\M{K}}$, and likewise consists of non-contiguous memory accesses. 
Naively loading $\hat{\M{V}}$ results in scattered memory instructions---for example, four separate 16-bit loads from non-adjacent addresses per thread, as shown on the top-right of \Cref{fig:regRemap}.

To mitigate this, we apply a similar register remapping to permute the column layout of $\hat{\M{V}}$ to increase the horizontal granularity of each thread's load (see bottom-right of \Cref{fig:regRemap}).
This results in a different layout of the output block, $\M{O}_i$. 
Since $\M{O}_i$ is stored in shared memory, we reverse this permutation when writing it back to HBM, so the final output layout matches the expected format.

We use the PTX \texttt{mma} interface rather than the CUDA \texttt{wmma} API. The key difference between the two is that \texttt{wmma} requires both input operands to reside in shared memory before being loaded into registers. In \kernel, the right-hand-side operands $\hat{\M{K}}_j^\Tra$ in SDDMM and $\hat{\M{V}}_j$ in SpMM are used only once per thread block and are not reused. Staging them in shared memory would introduce unnecessary data transfers and increase memory pressure without any performance benefit. With \texttt{mma}, we can load these operands directly from HBM into registers, reducing the number of memory operations and decreasing latency. 

\subsection{Mixed Precision and Stability}\label{sec:mixed-precision}

To optimize performance and memory footprint, \kernel employs a mixed-precision strategy. \Cref{tab:mixedp} summarizes the precision of the inputs, intermediate results, and output.
The input matrices $\M{Q}$, $\M{K}$, and $\M{V}$ are stored in \texttt{fp16} to reduce memory bandwidth requirements and leverage high-throughput \texttt{fp16} tensor core operations. Intermediate attention scores $\M{S}$, computed during SDDMM, are accumulated in \texttt{fp32} to minimize precision loss. Softmax is computed entirely in \texttt{fp32} for numerical stability. After softmax, the normalized scores $\M{E}$ are cast back to \texttt{fp16} before being stored in shared memory, as the subsequent SpMM accepts \texttt{fp16} inputs and produces the final output $\M{O}$ in \texttt{fp32}. This mixed-precision design balances performance and accuracy.

\begin{table}[h!]
    \centering
    \caption{Precision of inputs, intermediate results and output.}
    \label{tab:mixedp}
    \begin{tabular}{c|c|c|c|c|c|c}
        Matrix & $\M{Q}$ & $\M{K}$ & $\M{V}$ & $\M{S}$ & $\M{E}$ & $\M{O}$ \\\hline
        Precision & fp16 & fp16 & fp16 & fp32 & fp32 $\rightarrow$ fp16 & fp32 \\
    \end{tabular}
\end{table} 

\textbf{Softmax.}
Softmax is a key operation in attention, but it is susceptible to numerical overflow in low-precision formats. In its naive form, 
\begin{equation}
    \text{softmax}(\V{x}) = \frac{\exp(\V{x}_i)}{\sum_{j} \exp(\V{x}_j)}
\end{equation}
where the exponential may exceed the dynamic range of the floating point format. 
For example, the maximum value representable in \texttt{fp32} is approximately $e^{89}$. If any element in $\M{S}$ exceeds 89, its exponential becomes infinity, resulting in \texttt{NaN} values in the output. In \texttt{fp16}, the threshold is even lower—around $e^{11}$---making the problem more severe. 
These overflows not only corrupt inference results but also break differentiability during backpropagation.

Most attention implementations use the \emph{max-stabilized softmax} \cite{Goodfellow2016DeepLearning}, defined as:
\begin{equation}
    \text{softmax}(\V{x}) = \frac{\exp(\V{x}_i - \max(\V{x}))}{\sum_{j} \exp(\V{x}_j - \max(\V{x}))}
\end{equation}
which subtracts the row-wise maximum prior to exponentiation. 
Although the additional reduction (to compute $\max$) introduces synchronization overhead in GPU kernels, the gain in numerical stability is typically well worth the cost.

We implement the \emph{online softmax} algorithm \cite{dao2024flashattention2}, a blocked variant of the max-stabilized softmax. 
While online softmax can be less stable than the global variant, particularly with smaller block sizes \cite{golden2024flashattentionstable}, we find it to be a favorable trade-off for \kernel.
It significantly reduces memory consumption by avoiding the need to materialize the full attention score matrix and enables \kernel to scale to large graphs, as demonstrated in \Cref{sec:results}.

\section{Results}\label{sec:results}
We evaluate the performance of \kernel\ both as a standalone kernel and as the attention layers of a graph transformer model during inference on various real-world graphs of different sizes.

\subsection{Setup}\label{sec:results-setup}
\textbf{GPU Architecture.}
We perform experiments on NVIDIA A30 (Ampere) and H100 (Hopper) GPUs. The A30 has 56 SMs, each with 4 tensor cores and achieves up to 165 TFLOPs of FP16 tensor core throughput with 933 GB/s of DRAM bandwidth. 
The H100 has 132 SMs and 4 tensor cores per SM, delivering 990 TFLOPs of FP16 tensor core throughput and 4 TB/s of DRAM bandwidth.  
See NVIDIA datasheet for full architectural details~\cite{A30DataSheet,GH200DataSheet}.

\noindent\textbf{Datasets.}
We benchmark \kernel on a diverse set of graph datasets drawn from popular GNN benchmarks \cite{igbdatasets, snapnets, yang2016planetoid, zeng2020graphsaint, hamilton2018Reddit, NDR, weber2019Ellbitcoin, hu2020ogb}. \Cref{tab:datasets} summarizes their key properties. To characterize sparsity after the sparse matrix compaction described in \Cref{subsec:bsb}, we report two metrics: \texttt{\tcBlock/RW} and \texttt{nnz/\tcBlock}, assuming a \tcBlock\ size of $16 \times 8$. For both metrics, we include the coefficient of variation (CV = $
\sigma/\mu$), which quantifies irregularity. A high CV (\eg $\approx 1$) in \texttt{\tcBlock/\rowWindow} indicates a significant variation in workload per RW, which poses challenges for load balancing and warp-level parallelism. 

\begin{table}[ht]
    \centering
    \caption{Single Graph Datasets. Metrics shown are after sparse compaction with TC block size $16 \times 8$.}
    \small
    \setlength{\tabcolsep}{3.5pt} 
    \begin{tabular}{lrrrrrrr}
        \toprule
        \textbf{Name} & \textbf{Nodes} & \textbf{Edges}
        & \multicolumn{2}{c}{\textbf{\tcBlock/\rowWindow}} 
        & \multicolumn{2}{c}{\textbf{nnz/\tcBlock}} \\
        \cmidrule(lr){4-5} \cmidrule(lr){6-7}
        & & & \textbf{avg} & \textbf{CV} & \textbf{avg} & \textbf{CV} \\
        \midrule
        IGB-small   & $1$M  & $12.1$M & 24.4 & 0.25 & 7.9 & 0.11 \\
        IGB-medium   & $10$M & $120$M & 24.4 & 0.58 & 7.9 & 0.11 \\
        \midrule
        Amazon0505  & 410K  & 3.36M & 12.3 & 0.20 & 10.6 & 0.46 \\
        Com-Amazon  & 335K  & 926K & 6  & 0.61 & 7.5  & 0.22 \\
        Musae-github    & 38K   & 578K  & 29.4 & 1.34 & 8.3  & 0.15 \\
        Artist      & 51K   & 819K & 32   & 0.73 & 8  & 0.11 \\
        \midrule
        Pubmed      & 20K   & 89K & 9.3  & 0.45 & 7.7  & 0.18 \\
        Cora        & 2.7K  & 10.6K & 7.5  & 0.38 & 8.3  & 0.29 \\
        Citeseer    & 3.3K  & 9.2K & 5.8  & 0.31 & 7.7  & 0.24 \\
        \midrule
        AmazonProducts  & 1.57M & 264.3M & 330.5 & 1.22 & 8.2  & 0.07 \\
        Yelp        & 717K  & 14M & 39  & 1.28 & 8  & 0.09 \\
        \midrule
        Reddit      & 233K  & 114.9M & 477.2 & 1.35 & 16.5 & 0.95 \\
        \midrule
        Blog        & 89K   & 4.19M & 69  & 2.47 & 11 & 0.44 \\
        \midrule
        Elliptic    & 204K  & 234K & 2.5   & 0.57 & 7.5  & 0.45 \\
        \midrule
        Ogbn-products & 2.45M & 123.7M & 101.4 & 0.84 & 8  & 0.05 \\
        \bottomrule
    \end{tabular}
    \label{tab:datasets}
\end{table}

\begin{table*}[ht]
\small
\centering
\caption{Distribution of \tcBlock\ counts per \rowWindow\ across deciles. Each cell shows the min--max \tcBlock\ range in that decile.}
\label{tab:tcb_deciles}
\begin{tabular}{l*{11}{c}}
\toprule
\textbf{Dataset} & 
\textbf{decile size}&
\textbf{10\%}& 
\textbf{20\%}& 
\textbf{30\%}& 
\textbf{40\%}& 
\textbf{50\%}& 
\textbf{60\%}& 
\textbf{70\%}& 
\textbf{80\%}& 
\textbf{90\%}& 
\textbf{100\%}\\
\midrule
\textbf{Reddit}& 1456 & 4--46 & 46--88 & 88--135 & 135--190 & 190--265 & 265--367 & 367--503 & 503--718 & 718--1113 & 1114--9857 \\
\textbf{Yelp} & 4480 & 4--9 & 9--12 & 12--15 & 15--19 & 19--23 & 23--29 & 29--38 & 38--52 & 52--82 & 82--1000 \\
\textbf{Pubmed} & 123 & 1--5 & 5--6 & 6--7 & 7--8 & 8--9 & 9--10 & 10--11 & 11--12 & 12--14 & 14--43 \\
\textbf{Github} & 236 & 2--13 & 13--16 & 16--18 & 18--20 & 20--23 & 23--25 & 25--29 & 29--34 & 34--46 & 46--1191 \\
\bottomrule
\end{tabular}
\end{table*}

\Cref{tab:tcb_deciles} presents a more detailed breakdown of work imbalance for four representative graphs. We sort all \rowWindows by their \tcBlock\ count and group them into ten deciles. Each cell reports the minimum and maximum \tcBlock\ count within that decile. Graphs like Reddit exhibit a long tail: many sparse row windows, but some are extremely dense. Yelp and Github show similar irregularity, as reflected by their high CV values, making them useful for stress-testing load balance. In contrast, graphs like Pubmed have a more uniform distribution.

In addition to large single-graph datasets for node- and edge-level prediction tasks, many real-world applications--especially graph property prediction--process collections of small graphs (often fewer than 500 nodes). 
To improve GPU utilization, these graphs are batched together into a single larger graph. This batching introduces a unique sparsity pattern with many disconnected components.
We evaluate \kernel on batched graphs from two widely-used benchmarks: Long Range Graph Benchmark (LRGB) \cite{dwivedi2022LRGB} and Open Graph Benchmark (OGB) \cite{hu2020ogb}, with a batch size of 1024. 

\noindent\textbf{Baselines.}
We compare \kernel with the following competitive baselines for sparse attention and 3S:
\begin{itemize}
\item \textbf{DF-GNN}~\cite{liu2024dfgnn} is the state-of-the-art for the fused 3S kernel on CUDA cores in \texttt{fp32}, and includes a numerically stable softmax.
We evaluate two variants: \emph{tiling}, designed for larger graphs, and \emph{hyper}, optimized for small graphs. 
\item \textbf{FlashSparse}~\cite{shi2024flashsparse} represents the  state-of-the-art for SDDMM and SpMM as separate kernels on tensor cores with mixed precision (\texttt{fp16} + \texttt{fp32}). The original code implements a naive softmax, so we also benchmark a modified version with a numerically stable softmax for a fair comparison.
\item \textbf{PyTorch Geometric (PyG)}~\cite{FeyLenssen2019PyG} is a widely used GNN framework with a PyTorch backend.
\item \textbf{Deep Graph Library (DGL)}~\cite{wang2019deep} is another popular GNN framework. We include it in the end-to-end transformer evaluation, as the original Graph Transformer \cite{GraphTransformer} implementation is built on DGL.
\end{itemize}

\subsection{3S Kernel Performance}
\begin{figure*}[h]
    \centering
    \begin{subfigure}{\linewidth}
        \includegraphics[width=\linewidth]{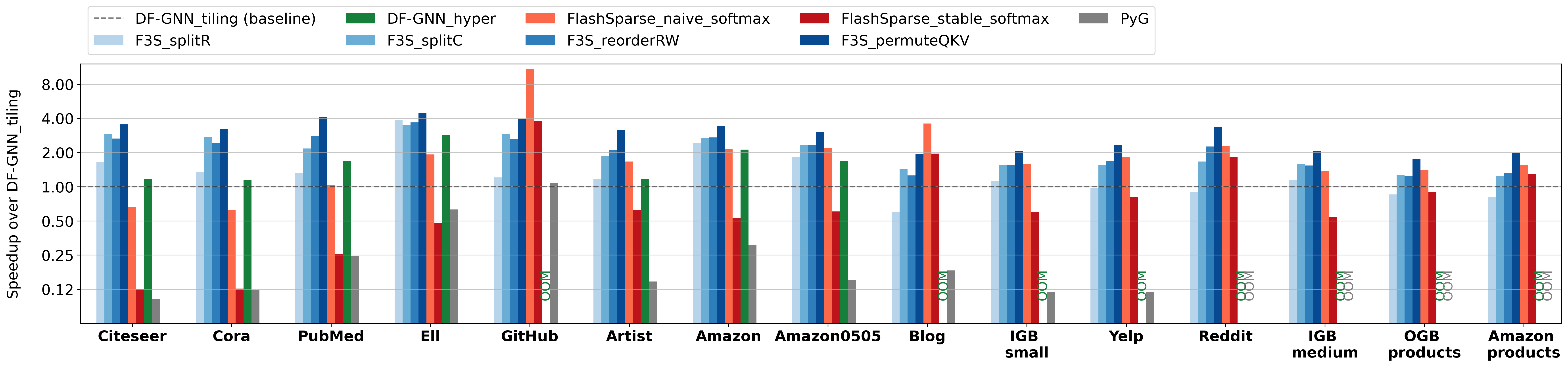}
        \caption{H100 GPU. Fused3S achieves $2.8\times$, $2.2\times$, $1.6\times$, $4.4\times$ and $14.7\times$ geometric mean speedup over DF-GNN\_tiling, DF-GNN\_hyper, FlashSparse\_naive\_softmax, FlashSparse\_stable\_softmax, and PyG respectively.}
    \end{subfigure}
    \begin{subfigure}{\linewidth}
        \includegraphics[width=\linewidth]{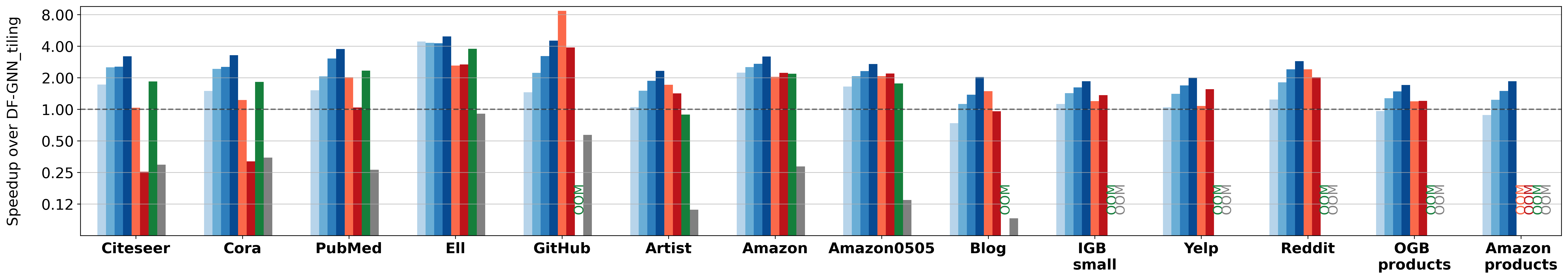}
        \caption{A30 GPU. Fused3S achieves $2.7\times$, $1.7\times$, $1.5\times$, $2.2\times$, and $12.3\times$ geometric mean speedup over DF-GNN\_tiling , DF-GNN\_hyper, FlashSparse\_naive\_softmax, FlashSparse\_stable\_softmax, and PyG respectively.}
    \end{subfigure}
    \caption{3S kernel performance on single graph datasets in Table \ref{tab:datasets}. Graphs are ordered by increasing number of edges (left to right). Y-axis is in log-scale.}\label{fig:3Sspeedup_fullgraphs}
\end{figure*}

\Cref{fig:3Sspeedup_fullgraphs} shows the 3S kernel performance on the single graph datasets listed in \Cref{tab:datasets}.
To summarize performance across the datasets, we report the geometric mean speedup computed as $\left(\prod_{d=1}^{D} s_d \right)^{\frac{1}{D}}$, where $s_d$ is the speedup on dataset $d$ and $D$ is the total number of datasets.

On the A30 and H100 GPUs, Fused3S consistently outperforms all baselines achieving geometric mean speedups of $1.5-12.3\times$ and $1.6-14.7\times$ respectively.  
DF-GNN\_hyper adopts a hybrid edge- and node-parallel strategy, partitioning non-zeros in SDDMM using edge-parallelism. This improves load balance and yields better performance than DF-GNN\_tiling on small graphs. However, it consumes significantly more memory since it stores entire rows of $\M{S}$ in shared memory. As a result, DF-GNN\_hyper fails on high-degree graphs such as Reddit, AmazonProducts, Ogbn-products, and IGB-medium.
In contrast, DF-GNN\_tiling which is based on node-parallel fusion uses less shared memory and is preferred for large graphs but suffers from load imbalance. 
FlashSparse outperforms its stable-softmax variant due to the additional synchronization required to compute row-wise max.
However, as discussed in \Cref{sec:mixed-precision}, the naive softmax is prone to overflow errors and is not recommended in practice.

Fused kernels avoid storing intermediate results between SDDMM, softmax, and SpMM, reducing memory pressure.
This is especially important on memory-constrained GPUs like the A30.
For example, on the AmazonProducts dataset with the most number of edges, both FlashSparse and PyG fail due to out-of-memory (OOM) errors caused by materializing the large $\M{S}$ matrix.
In contrast, Fused3S and DF-GNN\_tiling complete successfully. \kernel further benefits from mixed precision (\texttt{fp16}/\texttt{fp32}) execution, using less memory than DF-GNN, which runs entirely in \texttt{fp32}. On H100, \kernel remains the only kernel to run on the largest graphs tested (IGB-large and Ogbn-papers100M, results not shown).

On highly irregular graphs such as Blog, Yelp and Github, \kernel\ shows limited speedup. 
These datasets exhibit extreme variance in TCB counts per RW (see \Cref{tab:tcb_deciles}). For instance, in Github, 90\% of row windows have fewer than 46 TC blocks, while a few exceed 1000.
Even with row window reordering, such imbalance decreases compute and memory throughput. Assigning multiple thread blocks per row window could improve load balance. New GPU features such as \emph{thread block clusters} allow thread blocks within a cluster to synchronize in shared memory, which we plan to explore in future work. 

\begin{figure*}[h]
    \centering
    \begin{subfigure}{\linewidth}
    \includegraphics[width=\linewidth]{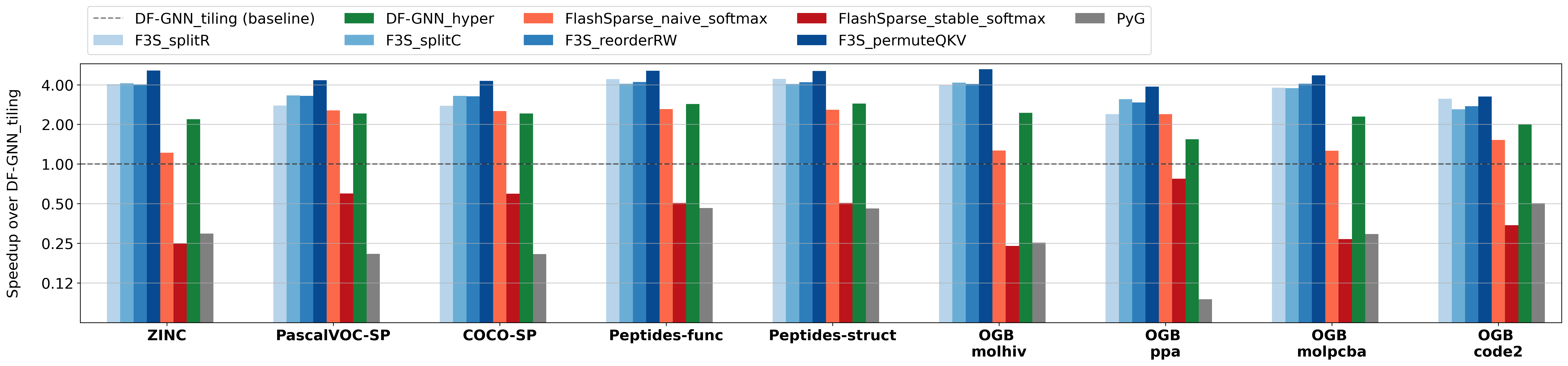}
    \caption{H100 GPU. Fused3S achieves $4.5\times$, $1.9\times$, $2.4\times$, $10.8\times$, and $16.3\times$ geometric mean speedup over DF-GNN\_tiling , DF-GNN\_hyper, FlashSparse\_naive\_softmax, FlashSparse\_stable\_softmax, and PyG respectively.}
    \end{subfigure}
        
    \begin{subfigure}{\linewidth}
    \includegraphics[width=\linewidth]{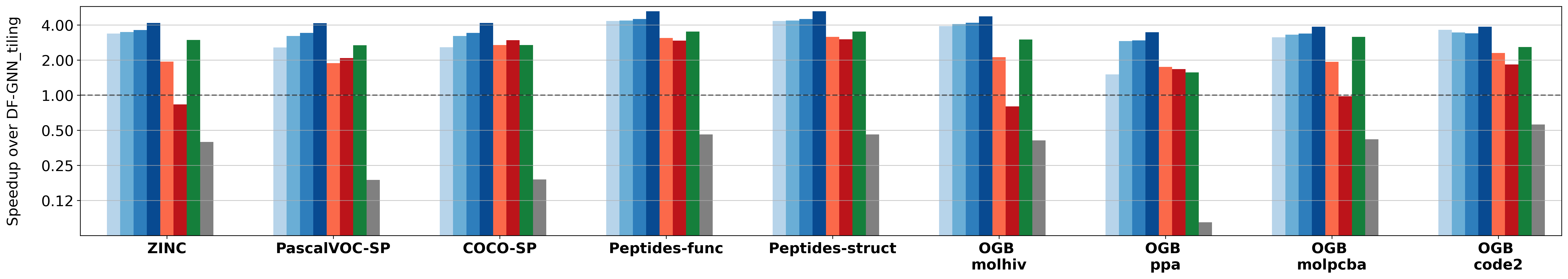}
    \caption{A30 GPU. Fused3S achieves $4.3\times$, $1.5\times$, $1.9\times$, $2.5\times$, and $14\times$ geometric mean speedup over DF-GNN\_tiling , DF-GNN\_hyper, FlashSparse\_naive\_softmax, FlashSparse\_stable\_softmax, and PyG respectively.}
    \end{subfigure}
    \caption{3S kernel performance on batched graphs from LRGB \cite{dwivedi2022LRGB} and OGB \cite{hu2020ogb}. Y-axis is in log-scale.}\label{fig:3Sspeedup_batchgraphs}
\end{figure*}

\Cref{fig:3Sspeedup_batchgraphs} shows the performance on batched graphs. On the A30 and H100 GPUs, Fused3S consistently outperforms all baselines achieving geometric mean speedups of $1.5-14\times$ and $1.9-16.3\times$ respectively.   
Batched graphs consist of disconnected components that exhibit more regular sparsity and clustering, which can be exploited to improve memory locality.
DF-GNN benefits from this naturally, whereas \kernel\ currently does not exploit component boundaries or subgraph-level structure. Incorporating such structure-aware optimizations is a promising direction for future work.

\subsection{\kernel Performance Breakdown}
We analyze the contribution of each kernel design decision in \kernel\ by incrementally enabling optimizations. Each variant builds upon the previous one, and their performance is shown in \Cref{fig:3Sspeedup_fullgraphs,fig:3Sspeedup_batchgraphs}.

\textbf{Warp partitioning.}
To evaluate the impact of warp partitioning strategies in \Cref{sec:warpPartitioning}, we compare two variants: \texttt{F3S\_splitR}, the combination of split-row SDDMM and split-column SpMM, and \texttt{F3S\_splitC}, the combination of split-column SDDMM and split-column SpMM.
On single graph datasets, \texttt{F3S\_splitC} achieves a geometric mean speedup of $1.5\times$ on both A30 and H100 GPUs.
The benefit is less pronounced on batched graphs, which tend to have lower degrees and fewer \tcBlocks per \rowWindow. As a result, the choice of warp partitioning has limited impact on overall performance.

\begin{figure}[h]
    \centering
    \begin{subfigure}{\linewidth}
        \includegraphics[width=\linewidth]{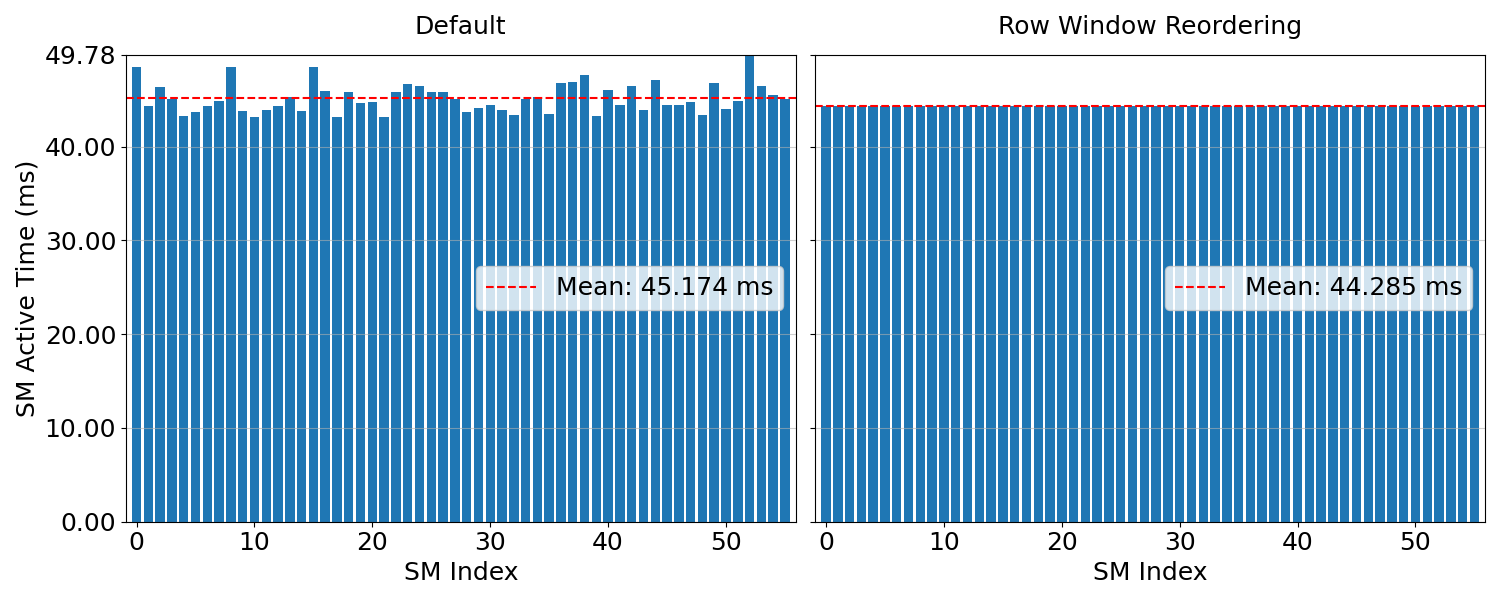}
        \caption{Reddit dataset}
    \end{subfigure}
    \begin{subfigure}{\linewidth}
        \includegraphics[width=\linewidth]{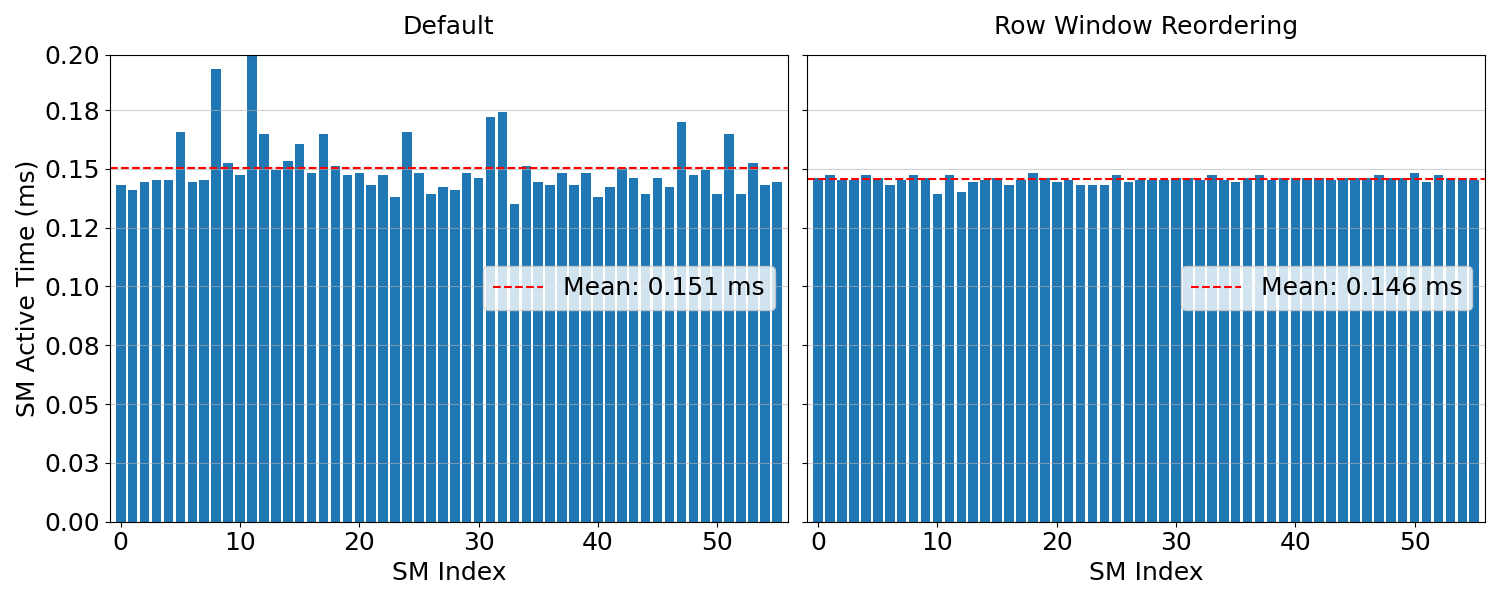}
        \caption{Pubmed dataset}
    \end{subfigure}
    \caption{Comparison of SM active time on A30 with (right) and without (left) row window reordering.}\label{fig:SM_Active_Comp}
\end{figure}

\textbf{Row window reordering.} 
We analyze the impact of sorting \rowWindows by their \tcBlock count. This optimization improves load balance by scheduling expensive row windows earlier in the kernel execution.
\Cref{fig:SM_Active_Comp} shows the active time of each of the 56 SMs on the A30 GPU, with and without reordering on two representative graphs (Reddit and Pubmed).
Without reordering, some SMs remain active for longer than others. 
On average, \texttt{F3S\_reorderRW} improves load balance and performance by $1.18\times$ over \texttt{F3S\_splitC} on about half of the single graph datasets.
However, the benefit depends on the graph structure.
When only a few \rowWindows contain many \tcBlocks while the rest are sparse (\eg Github and Blog), reordering offers limited gains (if any at all) as seen in \Cref{fig:3Sspeedup_fullgraphs,fig:3Sspeedup_batchgraphs}.

\textbf{Permuting $\M{Q}$, $\M{K}$, and $\M{V}$.}
We examine the effect of permuting the layout of $\M{Q}$, $\M{K}$, and $\M{V}$ as described in \Cref{sec:memAccess}. The \texttt{F3S\_permuteQKV} kernel applies this permutation on top of reordering and split-column partitioning. This optimization improves memory coalescing and instruction efficiency, achieving geometric mean speedups of $1.19-1.39\times$ on single graphs and $1.16-1.25\times$ on batched graphs.

\subsection{End-to-end Model Performance}
We evaluate the inference performance of the Graph Transformer (GT) model \cite{GraphTransformer} which comprises 10 transformer blocks, each with an attention layer, three feedforward layers, and two normalization layers. 
We replace the original attention kernel implemented in DGL \cite{wang2019deep} with four 3S variants: \kernel, DF-GNN's tiling and hyper kernels, and FlashSparse with naive softmax.

\begin{figure*}[h]
    \centering
        \begin{minipage}{1.0\linewidth}
            \centering
            \begin{subfigure}{\linewidth}
                \includegraphics[width=\linewidth]{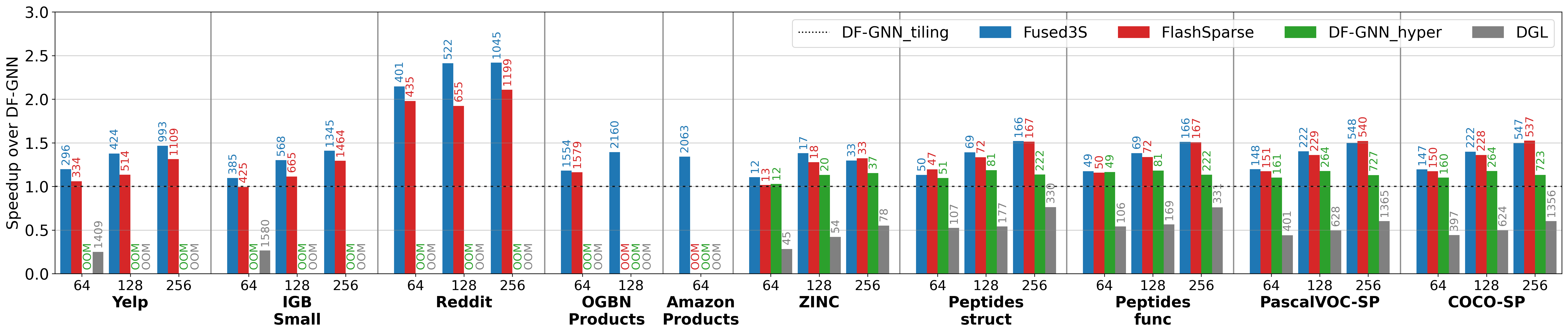}
                \caption{Performance on A30. Fused3S achieves $1.55\times$, $1.29\times$, $1.10\times$, and $3.08\times$ speedup over DF-GNN\_tiling, DF-GNN\_hyper, FlashSparse, and DGL respectively.}           
            \end{subfigure}
            \begin{subfigure}{\linewidth}
                \includegraphics[width=\linewidth]{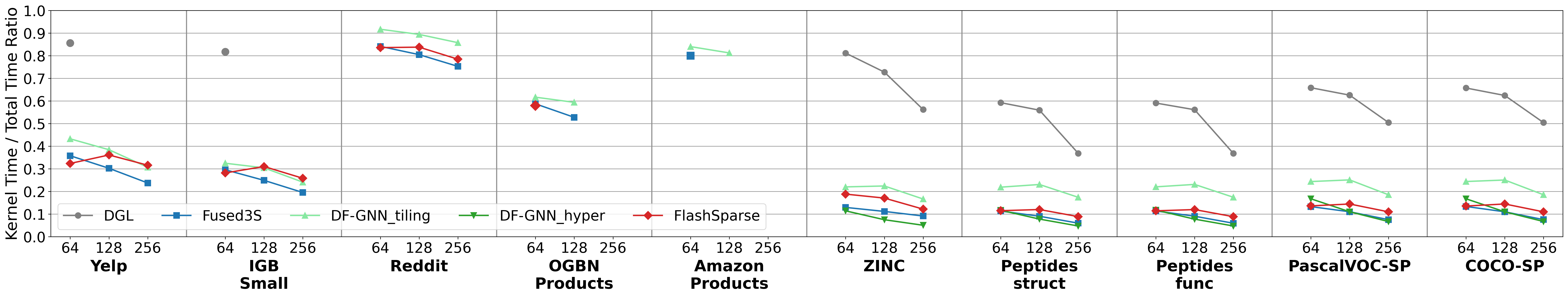}
                \caption{Fraction of inference time spent in attention kernels on A30.}
            \end{subfigure}
        \end{minipage}

        \begin{minipage}{1.0\linewidth}
            \centering
            \begin{subfigure}{\linewidth}
                \includegraphics[width=\linewidth]{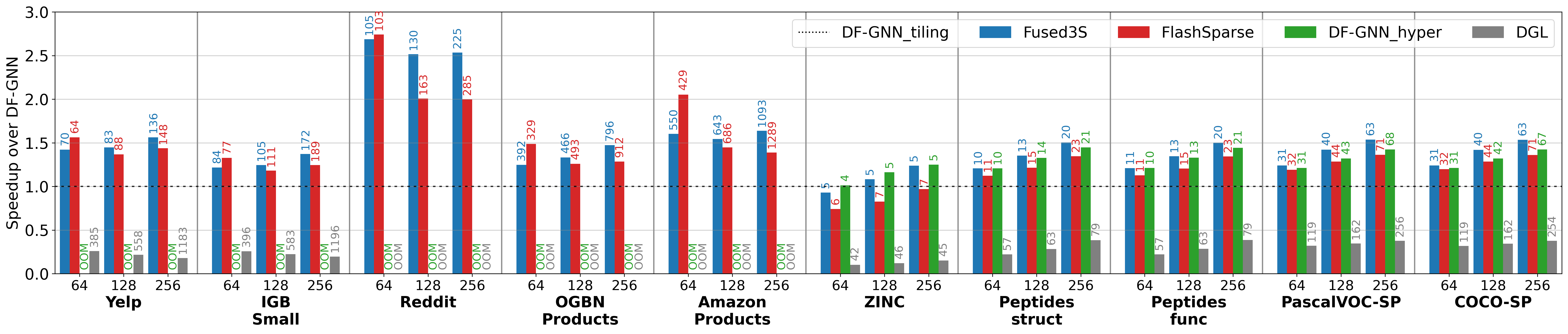}
                \caption{Performance on H100. Fused3S achieves $1.56\times$, $1.05\times$, $1.15\times$ and $5.36\times$ speedup over DF-GNN-tiling, DF-GNN-hyper, FlashSparse, and DGL respectively.}
            \end{subfigure}
            \begin{subfigure}{\linewidth}
                \includegraphics[width=\linewidth]{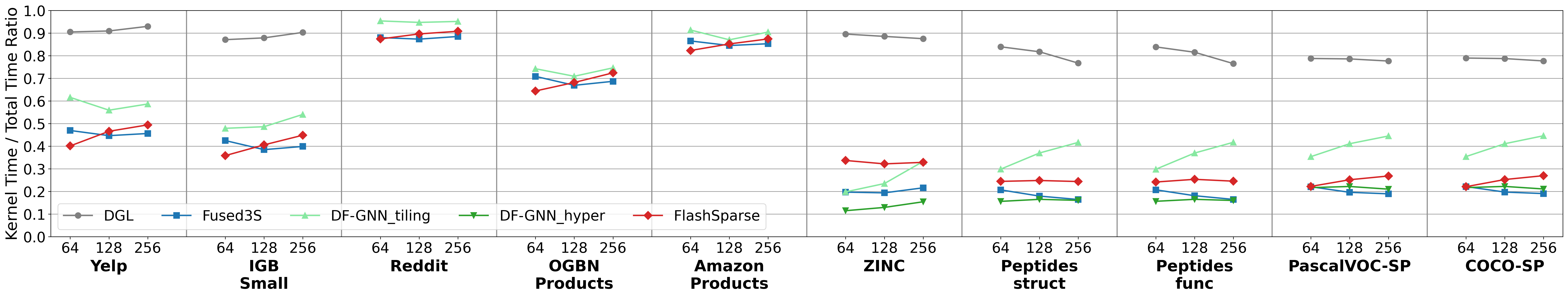}
                \caption{Fraction of inference time spent in attention kernels on H100.}
            \end{subfigure}
        \end{minipage}
    
    \caption{Graph Transformer inference performance with different 3S kernels. Missing bars indicate OOM. Labels on top of bars record the runtime in milliseconds.}
    \label{fig:gt_speedup}
\end{figure*}

\Cref{fig:gt_speedup} reports performance on five single graph and five batched graph datasets. For each dataset, we vary the embedding dimension $d \in \{64, 128, 256\}$ to assess sensitivity to model size. 
\kernel\ improves end-to-end inference time, achieving geometric mean speedups of $1.1-3.08\times$ and $1.05-5.36\times$ over the baselines on A30 and H100 respectively. 
As shown in \Cref{fig:gt_speedup}(b) and (d), the DGL baseline spends the majority of its inference time in the attention kernel. Replacing DGL with any optimized 3S kernel (including DF-GNN and FlashSparse) significantly reduces this bottleneck. 
As a result, attention accounts for a smaller fraction of the total inference time, partially amortizing kernel-level speedups, especially on smaller graphs.
The exceptions are larger graphs (Reddit, Ogbn-products, and AmazonProducts), where attention remains a bottleneck. 

Interestingly, the effect of increasing $d$ differs between A30 and H100.
On A30, increasing $d$ shifts the bottleneck toward the MLP layers, reducing the relative time spent in attention. In contrast, on H100, both MLP and attention layers scale efficiently, so attention remains a consistent or growing fraction of total time. This effect is particularly visible in DF-GNN\_hyper and FlashSparse, where shared memory pressure or lack of fusion limits the scalability of attention at higher $d$.

\section{Related Work}

\noindent\textbf{Sparse Matrix Computation on GPUs.}
Sparse matrix operations such as SpMM and SDDMM have received extensive attention due to their importance in GNNs, LLMs, and scientific computing. CUDA-core kernels such as Sputnik \cite{Sputnik} and RoDe \cite{RoDe} use 1D/2D tiling, offset alignment, memory coalescing, and load-balancing heuristics to target unstructured sparsity.
These approaches avoid preprocessing and operate directly on formats like CSR and COO.

With the growing adoption of tensor cores, recent efforts focus on enabling tensor core acceleration for sparse operations.  
TC-GNN \cite{TCGNN} proposes a TC-friendly format (TCF) that aligns sparsity patterns with MMA operand constraints; DTC-SpMM \cite{DTC-SpMM} extends this with ME-TCF and sparse double buffering to further reduce memory latency. Both DTC-SpMM and SMaT\cite{SMaT} use row reordering to increase the density of MMA tiles. FlashSparse~\cite{shi2024flashsparse} introduces separate optimized kernels for SDDMM and SpMM using the memory-efficient BCRS format and forming denser MMA tiles using 8$\times$1 vectors. Acc-SpMM \cite{zhao2024accspmm} proposes BitTCF, a compressed bitmask format for efficient tile decoding. 

Other designs focus on structured or semi-structured sparsity. JigSaw \cite{Jigsaw}, Flash-LLM \cite{Flash-LLM} and BSA-SpMM \cite{BSA-SpMM} focus on SpMM in transformer inference, where inputs are tall-skinny and sparsity is generated by weight pruning. 
TCA-SpMM \cite{TCA-SpMM} reshapes vector dot products into blocked matrix multiplications to improve TC utilization without preprocessing the sparse matrix into a different format. These techniques perform well under certain assumptions, but might not generalize to the irregular sparsity found in real-world graph data.

\noindent\textbf{Sparse Attention and Fused Kernels.}
Sparse attention typically involves three operations: SDDMM, softmax, and SpMM. Popular frameworks like DGL \cite{wang2019deep} and PyG \cite{FeyLenssen2019PyG} implement these as separate kernels and materializing intermediate outputs in memory. This results in significant memory traffic and kernel launch overhead.
DF-GNN \cite{liu2024dfgnn} is the first work to fuse all three operations into a single CUDA-core kernel with a numerically stable softmax.  It proposes two variants: tiling for large graphs and hyper for small graphs with high variance in node degree. However, DF-GNN executes entirely in \texttt{fp32}, uses CSR/COO/CSC formats, and does not target tensor cores. 

Sputnik, FlashSparse, and Magicube \cite{Magicube} also target sparse attention but do not fuse the softmax stage—intermediate results are materialized between SDDMM and SpMM. As a result, memory pressure remains high, and performance is bounded by inter-kernel synchronization.

\section{Conclusion}
\kernel\ is the first fully on-chip, fused sparse attention kernel designed for tensor cores. It introduces the BSB format, a block-aligned layout optimized for MMA operand shapes, and fuses SDDMM, softmax, and SpMM into a single TC-accelerated mixed-precision kernel. \kernel\ integrates  GPU optimizations including warp-level split-column partitioning, register remapping, and row window reordering to improve memory coalescing and address load imbalance under high sparsity. Experimental results show that \kernel achieves high performance on real-world graphs with unstructured sparsity--a use case not well supported by prior tensor core or fused sparse kernels.

Looking ahead, several directions offer potential for improving \kernel. 
Hopper's hardware features such as \texttt{fp8} compute and Tensor Memory Accelerator (TMA) could further reduce memory overhead and improve throughput. 
Alternative tile shapes enabled by lower precision, and operand reordering such as FlashSparse’s swap-and-transpose technique, may increase computational density. 
While this work focuses on the forward pass, extending the optimizations to the backward pass---which also involves SpMM and SDDMM operations in reverse order---is expected to yield similar performance improvements for training. Additionally, support for thread block clusters could enable synchronization across multiple thread blocks, unlocking finer-grained load balancing. Finally, adapting \kernel to better support the per-graph sparsity and block-disconnected structure of batched GNN datasets (\eg molecular graphs, abstract syntax trees, crystal graphs) may help bridge the gap between general sparse attention and multi-graph applications.

\begin{acks}
We thank the Research Cyberinfrastructure Center at UC Irvine for access to the GPUs on the HPC3 cluster.
We also thank Alex Danielian and Daniel Hsu for their assistance with kernel development and data preparation. 
\end{acks}

\clearpage
\bibliographystyle{ACM-Reference-Format}
\bibliography{ref}

\end{document}